# Orthogonal Frequency Division Multiplexing: An Overview

Kaur Inderjeet[1]
*Department of CSE[1]*
ITM, Gurgaon (INDIA)
inderjeetk@gmail.com

Sharma Kanchan[2]
*Department of ECE[2]*
IGIT, Delhi (INDIA)
sharma_kanchan@yahoo.com

Kulkarni.M[3]
*Department of ECE[3]*
DCE, Delhi (INDIA)
mkuldce@gmail.com

**Abstract**

*Orthogonal Frequency Division Multiplexing (OFDM) is a multi-carrier modulation scheme that provides efficient bandwidth utilization and robustness against time dispersive channels. This paper deals with the basic system model for OFDM based systems and with self-interference, or the corruption of desired signal by itself in OFDM systems. A simple transceiver based on OFDM modulation is presented. Important impairments in OFDM systems are mathematically analyzed.*

## 1. Introduction

The OFDM technology was first conceived in the 1960s and 1970s during research into minimizing **I**nter-**S**ymbol **I**nterference, or ISI, due to multipath. OFDM is a special form of **M**ulti **C**arrier **M**odulation (MCM) with densely spaced sub-carriers with overlapping spectra, thus allowing for multiple-access. MCM) is the principle of transmitting data by dividing the stream into several bit streams, each of which has a much lower bit rate, and by using these sub-streams to modulate several carriers. This technique is being investigated as the next generation transmission scheme for mobile wireless communications networks.

## 2. System Model

The Discrete Fourier Transform (DFT) of a discrete sequence f(n) of length N, F(k), is defined as [1],

$$F(k) = \frac{1}{N}\sum_{n=0}^{N-1} f(n)e^{-j\frac{2\Pi kn}{N}} \ldots\ldots(1)$$

and Inverse Discrete Fourier Transform (IDFT) as;

$$f(n) = \sum_{n=0}^{N-1} F(k)e^{j\frac{2\Pi kn}{N}} \ldots\ldots(2)$$

OFDM converts serial data stream into parallel blocks of size N, and uses IDFT to obtain OFDM signal. Time domain samples, then, can be calculated as

$$x(n) = IDFT\{X(k)\}$$
$$= \sum_{k=0}^{N-1} X(k)e^{j2\pi nk/N} \ldots 0 \le n \le N-1 \ldots(3)$$

where X(k) is the symbol transmitted on the kth sub-carrier and N is the number of sub-carriers. Symbols are obtained from the data bits using an M-ary modulation e.g. Binary Phase Shift Keying (BPSK), Quadrature Amplitude Modulation (QAM), etc. Time domain signal is cyclically extended to avoid Inter-symbol Interference (ISI) from previous symbol. The symbols X(k) are interpreted as frequency domain signal and samples x(n) are interpreted as time domain signal. Applying the central limit theorem, while assuming that N is sufficiently large, the x(n) are zero-mean complex-valued Gaussian distributed random variables. Power spectrum of OFDM signal with 64 sub-carriers is shown in Fig. 1. Symbols are mapped using Quadrature Phase Shift Keying (QPSK) modulation.

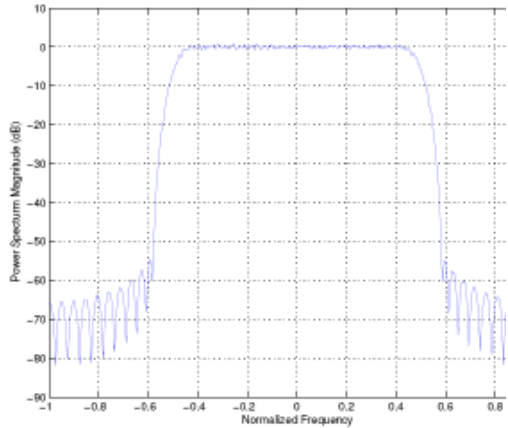

Fig 1. Power spectrum density of transmitted time domain OFDM signal.

To avoid difficulties in D/A and A/D converter offsets, and to avoid DC offset, the sub carrier falling at DC is not used as well. The power spectrum for such a system is shown in Fig. 2. Number of sub-carriers that are set to zero at the sides of the spectrum was 11.

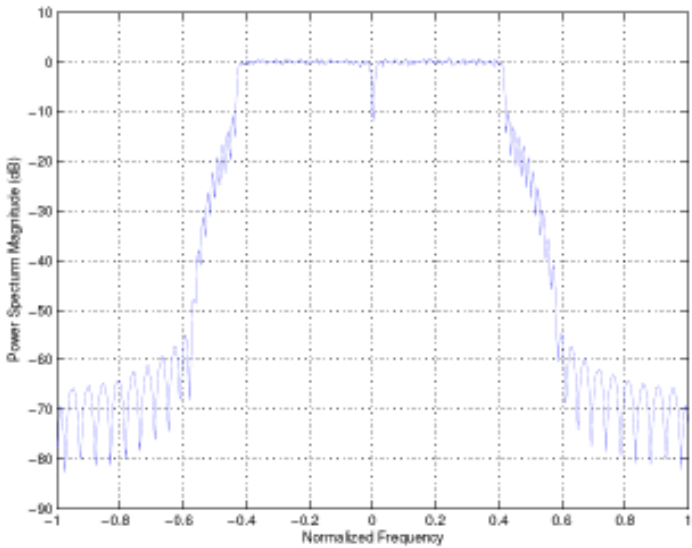

Fig 2. Power spectrum density of OFDM signal when the sub-carriers at the sides of the spectrum

## 2.1. Cyclic extension of OFDM symbol

Time domain OFDM signal is cyclically extended to mitigate the effect of time dispersion. The length of cyclic prefix has to exceed the maximum excess delay of the channel in order to avoid ISI [2,3]. The basic idea here is to replicate part of the OFDM time-domain symbol from back to the front to create a guard period. This is shown in the Fig. 3. This figure also shows how cyclic prefix prevents the ISI and as long as maximum excess delay ($\tau_{max}$) is smaller than the length of the cyclic extension (Tg), the distorted part of the signal will stay within the guard interval, which will be removed later at the transmitter. Therefore ISI will be prevented.

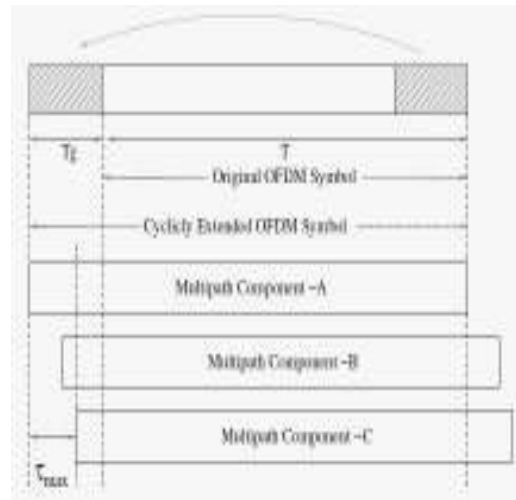

Fig 3. Illustration of cyclic prefix extension

The ratio of the guard interval to the useful symbol duration is application dependent. If this ratio is large, then the overhead will increase causing a decrease in the system throughput. A cyclic prefix is used for the guard time for the following reasons;

1. To maintain the receiver time synchronization; since a long silence can cause synchronization to be lost.
2. To convert the linear convolution of the signal and channel to a circular convolution and thereby causing the DFT of the circularly convolved signal and channel to simply be the product of their respective DFTs.
3. It is easy to implement in FPGAs.

### 2.2. Raised cosine guard period

The OFDM signal is made up of a series of IFFTs that are concatenated to each other. At each symbol boundary, there is a signal discontinuity due to the difference between the end of one symbol and the start of another one. These very fast transitions at the boundaries increase the side-lobe power. In order to smooth the transition between different transmitted OFDM symbols, windowing (Hamming, Hanning, Blackman, Raised Cosine etc.) is applied to each symbol.

### 2.3. Filtering

Filtering is applied both at the receiver and at the transmitter. At the transmitter, it is used to reduce the effect of side lobes of the sin shape in the OFDM symbol. This band pass filters removes some of the OFDM side-lobes. The amount of side-lobe removal depends on the sharpness of the filters used. In general digital filtering provides a much greater flexibility, accuracy and cut off rate than analog filters making them especially useful for band limiting of an OFDM signal [4].

### 2.4. Wireless channel

Communication channels introduce noise, fading, interference, and other distortions into the signals. The wireless channel and the impairments in the hardware of the receiver and transmitter introduce additive noise on the transmitted signal. The main sources of noise are thermal background noise, electrical noise in the receiver amplifiers, and interference. In addition to this, noise can also be generated internally to the communications system as a result of ISI, Inter-carrier Interference (ICI), and Inter-modulation Distortion (IMD) [4]. The noise due to these reasons decreases Signal-to-noise Ratio (SNR) resulting in an increase in the Bit Error Rate (BER).

Most of the noises from different sources in OFDM system can be modeled as Additive White Gaussian Noise (AWGN). AWGN has a uniform spectral density (making it white), and a Gaussian probability distribution.

### 2.5. A simple system

A block diagram of a basic OFDM system is given in Fig. 4. Usually raw data is coded and interleaved before modulation. In a multipath fading channel, all sub-carriers will have different attenuations. Some sub-carriers may even be completely lost because of deep fades. Therefore, the overall BER may be largely dominated by a few sub-carriers with the smallest amplitudes. To avoid this problem, channel coding can be used. By using coding, errors can be corrected up to a certain level depending on the code rate and type, and the channel. Interleaving is applied to randomize the occurrence of bit errors. Coded and interleaved data is then be mapped to the constellation points to obtain data symbols. These steps are represented by the first block of Fig. 4.

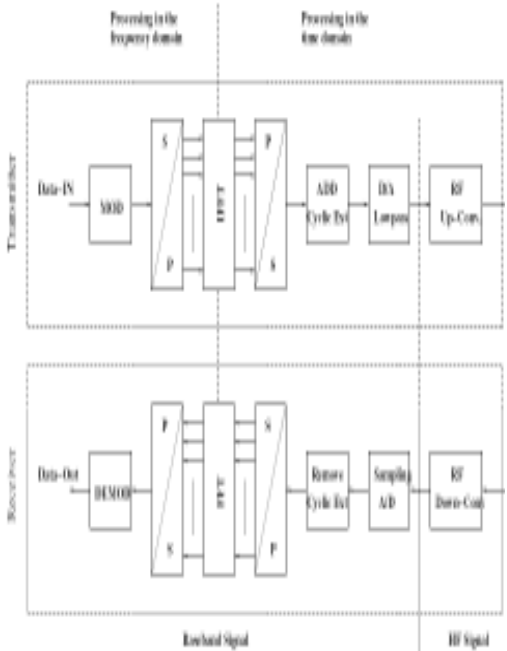

Fig 4. Block diagram of an OFDM transceiver.

The serial data symbols are then converted to parallel and Inverse Fast Fourier Transform (IFFT) is applied to these parallel blocks to obtain the time domain OFDM symbols. Later, these samples are cyclically extended as explained earlier and converted to analog signal and up-converted to the RF frequencies using mixers. The signal is then amplified by using a power amplifier (PA) and transmitted through antennas.

In the receiver side, the received signal is passed through a band-pass noise rejection filter and down converted to base band. After frequency and time synchronization, cyclic prefix is removed and the signal is transformed to the frequency domain using Fast Fourier Transform (FFT) operation. And finally, the symbols are demodulated, deinterleaved and decoded to obtain the transmitted information bits.

## 3. OFDM impairments

This section gives the main impairments that exist in OFDM systems with underlying mathematical details.

### 3.1. Frequency offset

Frequency offset is a critical factor in OFDM system design. It results in inter-carrier interference (ICI) and degrades the orthogonality of sub-carriers. Frequency errors will tend to occur from two main sources. These are local oscillator errors and common Doppler spread. Any difference between transmitter and receiver local oscillators will result in a frequency offset. This offset is usually compensated for by using adaptive frequency correction (AFC), however any residual (uncompensated) errors result in a degraded system performance.

The characteristics of ICI are similar to Gaussian noise; hence it leads to degradation of the SNR. The amount of degradation is proportional to the fractional frequency offset which is equal to the ratio of frequency offset to the carrier spacing. Frequency offset can be estimated by different methods e.g. using pilot symbols, the statistical redundancy in the received signal, or transmitted training sequences.

Assume that we have the symbols $X(k)$ to be transmitted using an OFDM system. These symbols are transformed to the time domain using IDFT as shown earlier in (3). This base band signal (OFDM symbol) is then up-converted to RF frequencies and transmitted over the wireless channel. In the receiver, the received signal is down-converted to base band. But, due to the frequency mismatch between the transmitter and receiver, the received signal has a frequency offset. This signal is denoted as y(n). The frequency offset is added to the OFDM symbol in the receiver. Finally, to recover the data symbols, DFT is applied to the OFDM symbol taking the signal back to frequency domain. Let Y (k) denote the recovered data symbols. This process is shown below:

$$= \frac{1}{N}\sum_{n=0}^{N-1}\left\{\sum_{m=0}^{N-1}X(m)e^{j\frac{2\pi n}{N}(m+\varepsilon)}\right\}e^{-j\frac{2\pi kn}{N}} \quad\ldots\ldots(10)$$

$$= \frac{1}{N}\sum_{n=0}^{N-1}\sum_{m=0}^{N-1}X(m)e^{j\frac{2\pi n}{N}(m-k+\varepsilon)} \quad\ldots\ldots(11)$$

$$= \frac{1}{N}\sum_{m=0}^{N-1}X(m)\left\{\sum_{n=0}^{N-1}e^{j\frac{2\pi n}{N}(m-k+\varepsilon)}\right\} \quad\ldots\ldots(12)$$

$$X(k)\xrightarrow{IDFT} x(n)\xrightarrow{frequency offset} y(n)\xrightarrow{DFT} Y(k)$$

### 3.2. Phase noise

Phase noise is introduced by local oscillator in any receiver and can be interpreted as a parasitic phase modulation in the oscillator's signal. Phase noise can be modeled as a zero mean random variable.

Let us apply the above operations to X(k) in order to get Y (k). First find x(n) using (2)

$$x(n) = IDFT\{X(k)\}\ldots\ldots\ldots\ldots(4)$$

$$= \sum_{k=0}^{N-1}X(k)e^{j2\pi nk/N}\ldots 0\le n\le N-1\ldots(5)$$

### 4. Peak-to-average power ratio

One of the major drawbacks of OFDM is its high Peak-to-average Power Ratio (PAPR). Superposition of a large number of sub-carrier signals results in a power density with Rayleigh distribution which has large fluctuations. OFDM transmitters therefore require power amplifiers with large linear range of operation which are expensive and inefficient. Any amplifier non-linearity causes signal distortion and inter-modulation products resulting in unwanted out-of-band power and higher BER [8]. The Analog to Digital converters and Digital to Analog converters are also required to have a wide dynamic range which increases complexity. Discrete-time PAPR of mth OFDM symbol $x_m$ is defined as [9]

The effect of frequency offset on x(n) will be a phase shift of $2\pi\varepsilon n/N$, where ε is the normalized frequency offset. Therefore;

$$y(n) = x(n)\times e^{j\frac{2\pi\varepsilon n}{N}}\ldots\ldots\ldots\ldots(6)$$

$$= \sum_{k=0}^{N-1}X(k)e^{j\frac{2\pi kn}{N}}\times e^{j\frac{2\pi\varepsilon n}{N}} \quad\ldots\ldots(7)$$

$$= \sum_{k=0}^{N-1}X(k)e^{j\frac{2\pi n}{N}(k+\varepsilon)} \quad\ldots\ldots\ldots(8)$$

Finally, we need to apply DFT to y(n) with a view toward recovering the symbols.

$$PAPR_m = \max_{0\le n\le N-1}|x_m(n)|^2 / E\{|x_m(n)|^2\}\ldots\ldots(13)$$

$$Y(k) = DFT\ (y(n))\ldots\ldots\ldots(9)$$

Although the PAPR is moderately high for OFDM, high magnitude peaks occur relatively rarely and most of the transmitted power is concentrated in signals of low amplitude, e.g. maximum

PAPR for an OFDM system with 32 carriers and QPSK modulation will be observed statistically only once in 3.7 million years if the duration of an OFDM symbol is 100μs [10]. Therefore, the statistical distribution of the PAPR should be taken into account.

Fig. 5 shows the probability that the magnitude of the discrete-time signal exceeds a threshold $x_0$ for different modulations. The number of sub-carriers was 128.

Applying the central limit theorem, while assuming that N is sufficiently large, x(n) is zero-mean complex-valued near Gaussian distributed random variables for all modulation options. Therefore, PAPR is independent of modulation used. This can be seen from Fig. 5.

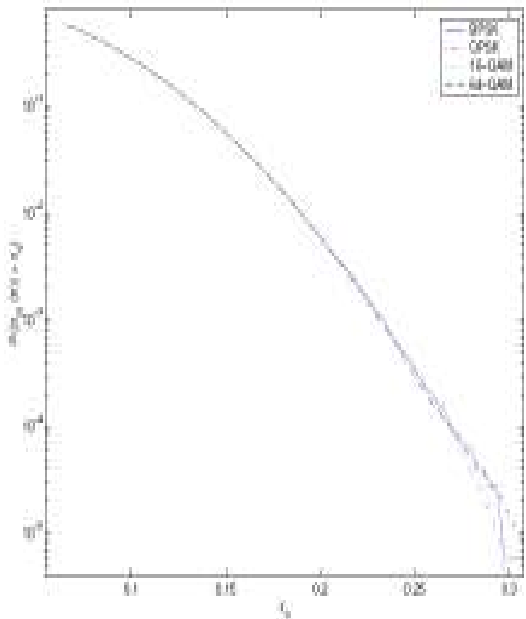

Fig 5. The probability that the magnitude of the discrete-time OFDM signal exceeds a threshold $x_0$ for different modulations.

One way to avoid non-linear distortion is to operate the amplifier in its linear region. Unfortunately such solution is not power efficient and thus not suitable for battery operated wireless communication applications. Minimizing the PAPR before power amplifier allows a higher average power to be transmitted for a fixed peak power, improving the overall signal to noise ratio at the receiver. It is therefore important to minimize the PAPR. The different approaches used to reduce PAPR of OFDM signals are clipping, scrambling, coding, phase optimization [8], tone reservation [10] and tone injection.

## 5. Conclusion

The demand for high data rate wireless communication has been increasing dramatically over the last decade. One way to transmit this high data rate information is to employ well known conventional single-carrier systems. Since the transmission bandwidth is much larger than the coherence bandwidth of the channel, highly complex equalizers are needed at the receiver for accurately recovering the transmitted information. Multi-carrier techniques can solve this problem significantly if designed properly. Optimal and efficient design leads to adaptive implementation of multi-carrier systems. Examples to adaptive implementation methods in multi-carrier systems include adaptation of cyclic prefix length, sub-carrier spacing etc. These techniques are often based on the channel statistics which need to be estimated.

In this paper, methods to estimate parameters for one of the most important statistics of the channel which provide information about the frequency selectivity have been studied.

These parameters can be used to change the length of cyclic prefix adaptively depending on the channel conditions. They can also be very useful for other transceiver adaptation techniques.

Current applications of Orthogonal Frequency Division Multiplexing (OFDM) do not require high mobility. For next generation applications, however, it is crucial to have systems that can tolerate high Doppler shifts caused by high mobile speeds. Current OFDM systems assume that the channel is time-invariant over OFDM symbol. As mobility increases, this assumption will not be valid anymore, and variations of the channel during the OFDM symbol period will cause ICI as explained. ICI due to time varying channel should be investigated further and effective channel estimation methods that are immune to ICI due to mobility should be developed to have OFDM ready for high mobility applications.

**References**
- [1] S. K. Mitra, Digital Signal Processing: A Computer-Based Approach, 2nd ed. New York, NY: McGraw-Hill, 2000.
- [2] W. Henkel, G. Taubock, P. Odling, P. Borjesson, and N. Petersson, "The cyclic prefix of OFDM/DMT - an analysis," in International Zurich Seminar on Broadband Communications. Access, Transmission, Networking., Zurich, Switzerland, Feb. 2002, p.1/3.
- [3] F. Tufvesson, "Design of wireless communication systems - issues on synchronization, channel estimation and multi-carrier systems," Ph.D. dissertation, Lund University, Aug. 2000.
- [4] E. P. Lawrey, "Adaptive techniques for multiuser OFDM," Ph.D. dissertation, James Cook University, Dec. 2001.
- [5] P. Dent, G. Bottomley, and T. Croft, "Jakes fading model revisited," IEE Electron. Lett., vol. 29, no. 13, pp. 1162-1163, June 1993.
- [6] P. H. Moose, "A technique for orthogonal frequency division multiplexing frequency offset correction," IEEE Trans. Commun., vol. 42, no. 10, pp. 2908-2914, Oct. 1994.
- [7] G. Armada, "Understanding the effects of phase noise in orthogonal frequency division multiplexing (OFDM)," IEEE Trans. Broadcast., vol. 47, no. 2, pp. 153-159, June 2002.
- [8] T. May and H. Rohling, "Reducing the peak-to-average power ratio in OFDM radio transmission systems," in Proc. IEEE Veh. Technol. Conf., vol. 3, Ottawa, Ont., Canada, May 1998, pp. 2474-2478.
- [9] S. Muller and J. Huber, "A comparison of peak power reduction schemes for OFDM," in Proc. IEEE Global Telecommunications Conf., vol. 1, Phoenix, AZ, USA, Nov. 1997, pp. 1-5.
- [10] S. Muller and J. Huber, "A comparison of peak power reduction schemes for OFDM," in Proc. IEEE Global Telecommunications Conf., vol. 1, Phoenix, AZ, USA, Nov. 1997, pp. 1-5.
- [11] H. Ochiai and H. Imai, "On the distribution of the peak-to-average power ratio in OFDM signals," IEEE Trans. Commun., vol. 49, no. 2, Feb. 2001.